\documentclass{conm-p-l}
\usepackage{epsfig}
\usepackage{epsf}
\usepackage{eepic}

\theoremstyle{definition}

\theoremstyle{remark}

\numberwithin{equation}{section}

\begin{document}

\title[Gap probabilities]{ On the gap probability generating function at the
spectrum edge in the case of orthogonal symmetry}

\author{P.J.~Forrester}

\curraddr{
 Department of Mathematics and Statistics, 
University of Melbourne, Victoria
3010, Australia}

\email{p.forrester@ms.unimelb.edu.au}

\thanks{Supported by the Australian Research Council}


\dedicatory{This paper is dedicated to Percy Deift on the occasion of his 60th birthday}


\begin{abstract}
The gap probability generating function has as its coefficients the probability
of an interval containing exactly $k$ eigenvalues. For scaled random matrices with orthogonal
symmetry, and the interval at the hard or soft spectrum edge, the
gap probability generating functions have the special property that  they can be evaluated
in terms of Painlev\'e transcendents. The derivation of these results
 makes use of formulas for the same generating function in certain scaled, 
 superimposed ensembles expressed in terms of its correlation functions.
It is shown that by a judicious choice of the superimposed ensembles, the scaled limit
necessary to derive these formulas can be rigorously justified by a straight forward
analysis.
\end{abstract}

\maketitle

\section{Introduction}
\subsection{An applied setting for gap probabilities}
The first use of random matrices to problems in theoretical physics was in relation
to the study of
the spectra of heavy nuclei (see \cite{Po65} for a collection of early
works on the subject). It was hypothesized that the highly excited energy levels of
heavy nuclei would have the same statistical properties as the the eigenvalues
from an ensemble of large random 
real symmetric matrices. More explicitly, the large
random real symmetric matrices were chosen from the Gaussian orthogonal ensemble
(GOE) in which each matrix $X$ occurs with probability density $e^{-{\rm Tr} X^2/2}$
(such an ensemble is invariant under the transformation $X \mapsto OXO^T$ where
$O$ is a real orthogonal matrix; this has the physical interpretation of there
being no preferential basis and explains too the adjective orthogonal in GOE).
To leading order matrices from the GOE have the support of their eigenvalues in
$[-\sqrt{2N}, \sqrt{2N}]$. The largest eigenvalue thus occurs in the
neighbourhood of $\sqrt{2 N}$, which is referred to as the soft edge, while the region away
from the edges (for example in the neighbourhood of the origin) is referred to as
the bulk. It is the statistical properties of the eigenvalues of large GOE
matrices in the bulk, scaled so that their mean spacing is unity, which are
compared against data from the spectra of heavy nuclei (with the latter also
scaled so that the mean spacing between consecutive levels is unity).

More recently problems from statistical physics have led to applications of
distributions at the soft edge of random matrix ensembles (i.e.~in the 
neighbourhood of the largest eigenvalue.) 
Here each of the three symmetry classes ---
orthogonal, unitary and symplectic --- are relevant to the applications.
In terms of Gaussian matrices, such symmetry classes are realized by the
probability density $e^{-\beta {\rm Tr} X^2/2}$ in which the Hermitian matrix
$X$ has real elements in the case of orthogonal symmetry $(\beta = 1$),
complex elements in the case of unitary symmetry $(\beta = 2)$ and real quaternion
elements represented as $2 \times 2$ matrices in the case of symplectic symmetry
$(\beta = 4)$. In the latter case the corresponding matrix $X$ has doubly
degenerate eigenvalues, and the convention is to count only one of the distinct
eigenvalues in the trace. For each of the three symmetry classes the corresponding 
eigenvalue probability density
function (p.d.f.) is of the form
\begin{equation}\label{2}
{1 \over C} \prod_{l=1}^N g_\beta (x_l) \prod_{1 \le j < k \le N}
| x_k - x_j |^\beta
\end{equation}
with $g(x) = e^{- \beta x^2}$ ($C$ denotes the normalization). Independent of $\beta$ 
the leading order support is on $[-\sqrt{2N}, \sqrt{2N}]$, and in the neighbourhood of 
the largest eigenvalue at $x = \sqrt{2N}$ (the soft edge) the eigenvalues have
spacing $O(1/N^{1/6})$. With $E_{N,\beta}(k;(s,\infty);g_\beta(x))$ denoting the
probability that exactly $k$ eigenvalues are in the interval $(s,\infty)$ of the
ensemble as specified by (\ref{2}) ($E_{N,\beta}$ is broadly referred to as a
gap probabilty), in keeping with these facts one expects
\begin{equation}\label{3}
\lim_{N \to \infty} E_{N,\beta} \Big ( k; ( \sqrt{2N} + c_\beta X / N^{1/6},
\infty) ; e^{- \beta x^2 / 2} \Big )
\end{equation}
to be a well defined order one quantity (here $c_\beta$ is an $N$ independent factor
chosen for convenience). Significantly, it is also expected that for all
$g_\beta(x)$ such that the eigenvalue support is to leading order a single interval
with right endpoint $a(N)$, there will be a scale $b(N)$ such that
\begin{equation}\label{4}
\lim_{N \to \infty} E_{N,\beta} \Big ( k; ( a(N) + b(N) X,
\infty) ; g_\beta(x) \Big )
\end{equation} 
exists and is equal to (\ref{3}). Such universality questions have been a major
theme of P.~Deift and collaborators (see \cite{De99} for a summary of this work 
up to 1999 relating to $\beta = 2$, and \cite{DG04,DG05,CDG06,DGKV06} 
for recent results on $\beta =1$ and 4).
In particular, with
$$
g_\beta(x) = x^a e^{-\beta x / 2}, \qquad (x > 0)
$$
which corresponds to the so called Laguerre ensemble, it is known from the rigorous work
of Johansson and Johnstone \cite{Jo99a,Jo01} that for $\beta = 1$ and 2
\begin{eqnarray*}
&& \lim_{N \to \infty} E_{N,\beta} (k; (\sqrt{2N} + X/ \sqrt{2} N^{1/6}, \infty);
e^{-\beta x^2} ) \nonumber \\
&& 
 =  \lim_{N \to \infty} E_{N,\beta} (k; (4 N + 2 (2N)^{1/3}, \infty) ;
x^a e^{- \beta x/2} ).
\end{eqnarray*}
This limiting probability is denoted $E_\beta^{\rm soft}(k;(s,\infty))$.

One class of problems in statistical physics giving rise to these probabilities relates to a
last passage percolation problem, originally formulated by Hammersley, and various
symmerizations due to Baik and Rains \cite{BR01a}. Here a unit
square, with bottom left corner at the origin,
contains points uniformly at random with a Poisson rate of intensity $z^2$
(this means that for $\delta$ small, each non-overlapping $\delta \times \delta$ subsquare
has probability $z^2 \delta^2$ of containing a point). Continuous piecewise linear
paths are formed from $(0,0)$ to $(1,1)$ by joining points in the square with segments of
positive slope. The length of a path is defined as the number of points it passes through,
and $l^U$ is used to denote the maximum of the length of all possible paths. The
limit theorem of Baik, Deift and Johansson \cite{BDJ98} tells us that
$$
\lim_{z \to \infty} {\rm Pr} \Big ( {l^U - 2 z \over z^{1/3} } \le y \Big ) =
E_2^{\rm soft}(0;(y,\infty)),
$$
so relating to the soft edge gap probability with
$\beta = 2$. In regards to this probability with $\beta = 1$, let
$l^S/2$ denote the maximum length of all paths going from $(0,0)$ to this
diagonal. A limit theorem of Baik and Rains \cite{BR01a} gives
$$
\lim_{z \to \infty} {\rm Pr} \Big ( {l^S - 2z \over z^{1/3} } \le y \Big ) =
E_1^{\rm soft}(0;(y,\infty)).
$$
Furthermore, with the points constrained to be symmetrical about the lower left to upper right
diagonal, Baik and Rains \cite{BR01a} proved an analogous limit theorem relating to
$E_4^{\rm soft}$.

\subsection{The gap probability generating function and Fredholm determinants}
A special feature of the gap probabilities $\{E_\beta^{\rm soft}(k;(s,\infty)) \}_{k=0,1,\dots}$
for the random matrix couplings $\beta = 1, 2$ and 4
is that they can be expressed in terms of Fredholm determinants and
Painlev\'e transcendents. How this comes about has different features for each of the three
$\beta$ values. However a common step is that one introduces the generating function
$$
E_{N,\beta}(J;g(x);\zeta) = \sum_{k=0}^\infty (1 - \xi)^k E_{N,\beta}(k;J;g_\beta(x))
$$
(because $ E_{N,\beta}(k;J;g_\beta(x)) = 0$ for $k>N$ the sum terminantes at $k=N$). 
About $\xi = 0$ this has the expansion
\begin{equation}\label{SS}
E_{N,\beta}(J;g(x);\zeta)   = 
 1 + \sum_{n=1}^\infty {(-\xi)^n \over n! }
\int_J dx_1 \cdots \int_J dx_n \, \rho_{(n)}(x_1,\dots,x_n)
\end{equation} 
where $\rho_{(n)}$ denotes the $n$-point correlation function for the point process specified by
(\ref{2}).

The case $\beta = 2$ is the simplest, because the p.d.f.~(\ref{2}) then specifies a determinantal
point process, which means that its $n$-point correlation is an $n \times n$ determinant. 
Explicitly, with $\{p_j(x) \}_{j=0,1,\dots}$ denoting the monic orthogonal polynomials
with respect to the weight function $g_2(x)$, and $(a,b)_2 := \int^\infty_{-\infty}
g_2(x) a(x) b(x) \, dx$, one has that (see e.g.~\cite{Fo02})
\begin{equation}\label{SS1}
\rho_{(n)}(x_1,\dots,x_n) = \det \Big [ K_N(x_i,x_j) \Big ]_{i,j=1,\dots,n}
\end{equation}
where
\begin{eqnarray}\label{SS2}
K_N(x,y) & = & (g_2(x) g_2(y) )^{1/2} \sum_{j=0}^{N-1}
{p_j(x) p_j(y) \over (p_j,p_j)_2} \nonumber \\
& = & {(g_2(x) g_2(y))^{1/2} \over (p_{N-1}, p_{N-1})_2 }
{p_N(x) p_{N-1}(y) - p_{N-1}(x) p_N(y) \over x - y},
\end{eqnarray}
with the final equality due to the Christoffel-Darboux summation formula.
Substituting (\ref{SS1}) in (\ref{SS}) one recognises the final expression as an expansion
of a Fredholm determinant (see \cite{WW65}), giving
\begin{equation}\label{SS3}
E_{N,2}(J;g_2(x);\xi) = \det ( {\mathbb I} - \xi K_{N,J} )
\end{equation}
where $K_{N,J}$ denotes the integral operator on $J$ with kernel (\ref{SS2}). With $g_2(x)$
corresponding to the Gaussian or Laguerre weights as introduced above, the soft edge scaling
limit is easy to perform rigorously \cite{Jo99a}, leading to the result \cite{Fo93a}
\begin{equation}\label{SS4}
E_{N,2}^{\rm soft}((s,\infty);\xi) = \det ( {\mathbb I} - \xi K_{(s,\infty)}^{\rm soft} )
\end{equation}
where $K_{(s,\infty)}^{\rm soft}$ is the integral operator on $(s,\infty)$ with kernel
\begin{equation}\label{1.8a}
K^{\rm soft}(x,y) = { {\rm Ai}(x) {\rm Ai}'(y) - {\rm Ai}(y) {\rm Ai}'(x) \over x - y}.
\end{equation}
This is the sought Fredholm determinant evaluation, and it in turn can be used 
\cite{TW94a,AV95,BD00} to deduce a Painlev\'e transcendent evaluation.

The situation at $\beta =1$ and 4 is more complex. The correlations are now quaternion
determinants rather than scalar determinants. In relation to the former, let $X$ be a
$2N \times 2N$ antisymmetric matrix, and set
$$
Z_{2N} := {\mathbb I}_N \otimes
\left [ \begin{array}{cc} 0 & -1 \\ 1 & 0 \end{array} \right ].
$$
With the quaternion dual of a $2N \times 2N$ matrix specified by
$$
A^D = Z_{2N} A^T Z_{2N}^{-1}
$$
one notes that matrices of the form $X Z_{2N}$ are self quaternion dual. The quaternion
determinant qdet of such matrices has the key property \cite{Dy72} that
$$
{\rm qdet} \, XZ_{2N} = {\rm Pf} \, X
$$
where Pf denotes the Pfaffian. As a consequence, for $A$ self quaternion dual, the
quaternion determinant and ordinary determinant are related by
\begin{equation}\label{qA}
({\rm qdet} \, A)^2 = \det A.
\end{equation}
Analogous to (\ref{SS3}), one has that if, as is the case at $\beta = 1$ and 4 (see e.g.~\cite{Fo02}),
$$
\rho_{(n)}(x_1,\dots,x_n) = \det \Big [ \tilde{K}_N(x_i,x_j) \Big ]_{i,j=1,\dots,n}
$$
where $\tilde{K}_N$ is a $2 \times 2$ matrix such that 
$[ \tilde{K}_N(x_i,x_j)]_{i,j=1,\dots,n}$ is self quaternion dual, then
$$
E_{N,\beta}(J;g_\beta(x);\xi) = {\rm qdet} ( \mathbb I - \xi  \tilde{K}_{N,J}).
$$
Here $\tilde{K}_{N,J}$ is a $2 \times 2$ matrix integral operator on the interval $J$, and
$\tilde{K}_{N,J}$ has kernel $\tilde{K}_N(x,y)$. Squaring both sides and applying (\ref{qA})
shows
\begin{equation}\label{qA1}
\Big ( E_{N,\beta}(J;g_\beta(x);\xi) \Big )^2 =  {\rm det} ( \mathbb I - \xi  \tilde{K}_{N,J}).
\end{equation}

The formula (\ref{qA1}) with $\xi = 1$ is taken as the starting point of the analysis in
\cite{TW96} leading to Painlev\'e evaluations of $E_1^{\rm soft}(0;(s,\infty))$ and
$E_4^{\rm soft}(0;(s,\infty))$. This strategy is used in \cite{Di05a} to give the analogous
formulas for the corresponding generating functions. However the details of these calculations
are very technical. Fortunately there is another approach to the problem introduced by the
present author in \cite{Fo99b}, and to be further developed herein.

In this approach, instead of working with the soft edge scaled limit of
(\ref{qA1}), and thereby involving a matrix integral operator, the first step is to derive
the alternative expression
\begin{equation}\label{qA2}
\Big ( E_1^{\rm soft}(0;(s,\infty) \Big )^2 = E_2^{\rm soft}(0;(s,\infty))
\Big ( 1 - \int_s^\infty (\mathbb I
 - K_{(s,\infty)}^{\rm soft} )^{-1} A^{\rm s}[y] B^{\rm s}(y) \, dy \Big )
\end{equation}
where $K_{(s,\infty)}^{\rm soft} $ is as in (\ref{SS4}), $A^{\rm s}$ is the operator which multiplies by
Ai$(x)$, while $B^{\rm s}$ is the integral operator with kernel $\int_0^\infty {\rm Ai}(y-v) \, dv$.
Starting from (\ref{qA2}) the Painlev\'e expression can be deduced in a page or two of working
(assuming knowledge of some identities from \cite{TW94a}).  Another significant feature of
(\ref{qA2}) is that it has been taken as the starting point of the proof of the Fredholm
determinant formula \cite{Sa05, FS05}
$$
 E_1^{\rm soft}(0;(s,\infty) = \det (\mathbb I - V_{(0,\infty)}^{\rm soft} )
$$
where $V^{\rm soft}(x,y)$ is the integral operator on $(0,\infty)$ with kernel
$V^{\rm soft}(x,y) = {\rm Ai}(x+y+s)$.

To extend the approach of \cite{Fo99b} to the Painlev\'e evaluation of the generating function
$E_1^{\rm soft}((s,\infty);\xi)$ \cite{Fo06c}, a key identity is the formula
\begin{equation}\label{qA3}
E^{{\rm odd(OEsoft)^2}}((s,\infty);\xi) =
\det ( \mathbb I - \xi K_{(s,\infty)}^{\rm soft})
\Big ( 1 - \xi \int_s^\infty [ (\mathbb I - \xi K_{(s,\infty)}^{\rm soft})^{-1} A^{\rm s}]
(y) B^{\rm s}(y) \, dy \Big )
\end{equation}
which for $\xi = 1$ is equivalent to (\ref{qA2}). Here, with the ensemble (\ref{2}) for $\beta = 1$
referred to as OE${}_N(g_1(x))$ (here the ``O'' denotes the underlying orthogonal symmetry),
${\rm OE}_N(g_1(x)) \cup {\rm OE}_N(g_1(x))$ denoting the superposition of two independent
such ensembles, and the operation ``odd'' referring to observing only each odd labelled
eigenvalue in the superposition as counted from the right most eigenvalue (i.e.~the soft
edge), odd(OEsoft)${}^2$ refers to the soft edge scaling limit of the ensemble
\begin{equation}\label{qA3a}
{\rm odd} \Big ( {\rm OE}_N(e^{-x^2}) \cup  {\rm OE}_N(e^{-x^2}) \Big ).
\end{equation}

Another advantage of the approach of \cite{Fo99b} is that the Painlev\'e evaluations for
$\beta = 4$ are deduced as a corollary of those at $\beta = 1$ and $\beta = 2$. This is
possible because of inter-relations between the gap probabilities for the three symmetry
types \cite{FR01}.

The primary motivating factor in seeking an alternative approach to the derivation of the
Painlev\'e evaluations of the $\beta = 1$ and 4 soft edge gap probabilities was to calculate
analogous formulas for the hard edge. The hard edge refers to the neighbourhood of the origin
when $g_\beta(x) \sim x^a$ as $x \to 0^+$, $g_\beta(x) = 0$ for $x<0$. In the case $g_\beta(x) =
x^a e^{-\beta x/2}$, $x>0$, the appropriate hard edge scaling is $x \mapsto X/4N$, and working
based on superimposed ensembles can be carried through, leading to the sought Painlev\'e evaluations
\cite{Fo99b,Fo06c}.
More explicitly, let 
\begin{equation}\label{qA3b}
{\rm odd} \Big ( {\rm OE}_N(x^{(a-1)/2} e^{-x/2}) \cup   {\rm OE}_N(x^{(a-1)/2} e^{-x/2}) \Big )
=: {\rm even}({\rm LOE}_N)^2
\end{equation}
denote the joint distribution of all odd labelled eigenvalues in the superposition
(labelled from the hard edge at $x=0$). Let odd(OEhard)${}^2$ refer to the hard edge scaling of
this joint distribution. The identity which plays the role of (\ref{qA3}) at the hard edge is
\begin{eqnarray}\label{qA4}
&& E^{{\rm odd(OEhard)}^2}((0,s);\xi;\alpha) \Big |_{\alpha = (a-1)/2} \nonumber \\
&& \qquad =
\det ( \mathbb I - \xi K_{(0,s)}^{\rm hard} ) \Big ( 
1 - \xi \int_s^\infty [ (\mathbb I - \xi K_{(0,s)}^{\rm hard})^{-1} A^{\rm h}]
(y) B^{\rm h}(y) \, dy \Big ).
\end{eqnarray}
Here the argument $\alpha$ refers to the hard edge singularity $x^\alpha$,
$K_{(0,s)}^{\rm hard}$ is the integral operator on $(0,s)$ with kernel
\begin{equation}\label{1.15a}
K^{\rm hard} (x,y)= {J_a(\sqrt{x}) \sqrt{y} J_a'(\sqrt{y}) - \sqrt{x} J_a'(\sqrt{x})J_a(\sqrt{y})
\over 2(x - y)}
\end{equation}
while $A^{\rm h}(x,y)$ is the operator which multiplies by $J_a(\sqrt{x})$ and $B^{\rm h}$ is the
integral operator on $(0,s)$ with kernel ${1 \over 2 \sqrt{y}} \int_y^\infty J_a(t) \, dt$.

\subsection{Aim of the paper}
In this paper we will reconsider the derivation of (\ref{qA3}) and (\ref{qA4}), 
which in the unpublished work \cite{Fo99b} is not rigorous. We will
begin by recalling and giving a critique of the strategy used in \cite{Fo99b}. Then we will
proceed to present a rigorous strategy based on different ensembles than those used in
(\ref{qA3}) and (\ref{qA4}) to scale to  odd(OEsoft)${}^2$ and  odd(OEhard)${}^2$
respectively. These ensembles have the advantage that the closed form expressions for the
correlation functions are simpler than those of the original ensembles. This makes the analysis
of the scaling of the corresponding gap probabilities much simpler.

\section{Gap probability generating function for superimposed ensembles}
\subsection{Review and critique of the original calculation}
To study the derivation of (\ref{qA3}) and (\ref{qA4}) one first notes that for an
integral operator $\mathbb I + C \otimes D$, the fact that $C \otimes D$ is of rank 1 gives
that
$$
\det ( \mathbb I + C \otimes D ) = 1 + \int_{-\infty}^\infty C(y) D(y) \, dy.
$$
It follows that (\ref{qA3}) and (\ref{qA4}) can be rewritten
\begin{eqnarray}
E^{{\rm odd(OEsoft)^2}}((s,\infty);\xi) & = &
\det \Big ( \mathbb I - \xi \Big ( \tilde{K}_{(s,\infty)}^{\rm soft}) +
\tilde{A}^{\rm s} \otimes \tilde{B}^{\rm s} \Big ) \Big ) \label{b1} \\
E^{{\rm odd(OEhard)^2}}((0,s);\xi;(a-1)/2) & = &
\det \Big ( \mathbb I - \xi \Big ( \tilde{K}_{(0,s)}^{\rm hard}) +
\tilde{A}^{\rm h} \otimes \tilde{B}^{\rm h} \Big ) \Big ). \label{b2}
\end{eqnarray}
Here the tilde on the operators indicates that the kernels have been multiplied by the
gauge factor $(1/\tilde{a}(x)) \tilde{a}(y)$ for $\tilde{a}$ decaying sufficiently rapidly
so that the integral operators inside the determinants are trace class. This latter
technicality can be avoided by expanding (\ref{b1}) and (\ref{b2}) according to the right hand
side of (\ref{SS}) to obtain
\begin{eqnarray}\label{b1a}
&& E^{{\rm odd(OEsoft)}^2}((s,\infty);\xi) \nonumber
\\
&& \quad = 1 + \sum_{n=1}^\infty
{(-\xi)^n \over n!} \int_s^\infty dx_1 \cdots  \int_s^\infty dx_n \,
\rho_{(n)}^{{\rm odd(OEsoft)}^2}(x_1,\dots,x_n)
\end{eqnarray}
with
\begin{equation}\label{2.2a}
\rho_{(n)}^{{\rm odd(OEsoft)}^2}(x_1,\dots,x_n)
= \det \Big [ K^{\rm soft}(x_j,x_k) + {\rm Ai}(x_j) \int_0^\infty {\rm Ai}(x_k - v) \, dv
\Big ]_{j,k=1,\dots,n}
\end{equation}
and
\begin{eqnarray}\label{b1b}
&&E^{{\rm odd(OEhard)}^2}((0,s);\xi;(a-1)/2) \nonumber \\
&& \quad = 1 + \sum_{n=1}^\infty
{(-\xi)^n \over n!} \int_0^s dx_1 \cdots  \int_0^s dx_n \,
\rho_{(n)}^{{\rm odd(OEhard)}^2}(x_1,\dots,x_n)
\end{eqnarray}
with
\begin{equation}\label{2.3a}
\rho_{(n)}^{{\rm odd(OEhard)}^2}(x_1,\dots,x_n)
= \det \Big [ K^{\rm hard}(x_j,x_k) + {J_a(\sqrt{x_j}) \over 2 \sqrt{x_k} }
\int_{\sqrt{x_k}}^\infty J_a(t) \, dt \Big ]_{j,k=1,\dots,n}. 
\end{equation}

The derivation of (\ref{b1a}) given in \cite{Fo99b} took as its starting point an explicit form
of the $n$-point correlations
\begin{eqnarray}\label{SN}
&& \rho_{(n)}^{{\rm odd(GOE}{}_N)^2}(x_1,\dots,x_n) = \det \Big [ K_{N-1}(x_j,x_k) \nonumber \\
&& \qquad  + e^{-x_j^2/2} 2^{-(N-1)} H_{N-1}(x_j)
\Big ( A_1(x_k) + A_2(x_k) \Big ) \Big ]_{j,k=1,\dots,n} 
\end{eqnarray}
where $K_{N-1}$ is specified by (\ref{SS2}) with $g_2(x) = e^{-x^2}$, and
\begin{eqnarray}\label{SMP}
A_1(y) & = & {e^{-y^2/2} \over (N/2 - 1)!} \sum_{\nu = 0}^\infty
{(N/2 - 1 + \nu)! \over (N-1 + 2 \nu)! } H_{N-1 + 2 \nu} (y) \nonumber \\
A_2(y)  & = & { \pi^{1/2} e^{-y^2/2} \over (N/2 - 1)! }
\sum_{l=0}^\infty {1 \over 2^{N+ 2l} (N/2 + l)! } H_{N+2l}(y)
\end{eqnarray}
(in these formulas it is assumed that $N$ is even). The task now is to show that with $J = (s,\infty)$
and the soft edge scale $s \mapsto \sqrt{2N} + s/2^{1/2} N^{1/6}$, the large $N$ limit of
(\ref{SS}) with $\rho_{(2)}$ given by (\ref{SN}) is equal to (\ref{b1a}). We know from
\cite{So01} (see \cite{BF03} for a restatement of this) that for this task it is sufficient to show
\begin{eqnarray}\label{G1}
&& \lim_{N \to \infty} \int_{\sqrt{2N} + s/2^{1/2} N^{1/6}}^\infty dx_1 \cdots
 \int_{\sqrt{2N} + s/2^{1/2} N^{1/6}}^\infty dx_n \,
\rho_{(n)}^{{\rm odd(GOE}{}_N)^2}(x_1,\dots,x_n) 
\nonumber 
\\ && \qquad
=
\int_s^\infty dx_1 \cdots \int_s^\infty dx_n \, \rho_{(n)}^{{\rm odd(OEsoft})^2}(x_1,\dots,x_n).
\end{eqnarray}

The starting point of the derivation of (\ref{b1b}) given in \cite{Fo99b} is very similar. In the 
notation of (\ref{qA3b}), the analogue of (\ref{SN}) is
\begin{eqnarray}\label{SNp}
&& \rho_{(n)}^{{\rm odd(LOE}{}_N)^2}(x_1,\dots,x_n) 
 = \det \Big [ K_{N-1}(x_j,x_k) + (g_2(x_j))^{1/2} L_{N-1}^a(x_j)  \nonumber \\
&& \quad \times {(N-1)! \over 2^{N-2} ((N-2)/2)! (a/2 + (N-2)/2)! }
\Big ( B_1(x_k) + B_2(x_k) \Big ) \Big ]_{j,k=1,\dots,n}.
\end{eqnarray}
Here $K_{N-1}(x,y)$ refers to (\ref{SS2}) with $g_2(x) = x^a e^{-x}$ and (assuming $N$ is even)
\begin{gather}\label{SN1}
B_1(y)  =  \sum_{\nu = (N-2)/2}^\infty
{2^{2\nu} (a/2 + \nu)! \nu! \over (a + 2 \nu + 1)!} ( g_2(y))^{1/2} L_{2 \nu + 1}^a(y) \nonumber \\
B_2(y)  = 2^{a-1} {((a-1)/2)!^2 (a/2)!^2 \over a!^2 }
\sum_{l=N/2}^\infty {(2l)! \over 2^{2l} l! (a/2 + l)! } (g_2(y) )^{1/2} L_{2l}(y). 
\end{gather}
Here the task is to show that with $J=(0,s)$, and the hard edge scale $s \mapsto s/4N$, the
large $N$ limit of (\ref{SS}) with $\rho_{(2)}$ given by (\ref{SN1}) is equal to (\ref{b1b}).
For this, according to \cite{So01}, it is sufficient to show
\begin{eqnarray}\label{G2}
&& \lim_{N \to \infty} \int_0^{s/4N} dx_1 \cdots
 \int_0^{s/4N} dx_n \,
\rho_{(n)}^{{\rm odd(LOE}{}_N)^2}(x_1,\dots,x_n) \nonumber \\
&& \quad =
\int_0^s dx_1 \cdots \int_0^s dx_n \, \rho_{(n)}^{{\rm odd(OEhard})^2}(x_1,\dots,x_n).
\end{eqnarray}

A mechanism for the validity of (\ref{G1}) and (\ref{G2}) is the uniform estimates 
\begin{eqnarray}\label{G3a}
&&\Big ( {1 \over \sqrt{2} N^{1/6} }\Big )^n
\rho_{(n)}^{{\rm odd(LOE}{}_N)^2} (\sqrt{2N} + x_1/2^{1/2} N^{1/6}, \dots, 
\sqrt{2N} + x_n/2^{1/2} N^{1/6}) \nonumber \\
&& \qquad =
\rho_{(n)}^{{\rm odd(OEsoft)}^2}(x_1,\dots,x_n) + o(1) R_n^{\rm s}(x_1,\dots,x_n)
\end{eqnarray}
and
\begin{eqnarray}\label{G3b}
&&\Big ( {1 \over 4 N} \Big )^n
\rho_{(n)}^{{\rm odd(LOE}{}_N)^2} (x_1/4N, \dots, x_n/4N) \nonumber \\ && \quad =
\rho_{(n)}^{{\rm odd(OEhard)}^2}(x_1,\dots,x_n) + o(1) R_n^{\rm h}(x_1,\dots,x_n)
\end{eqnarray}
where $o(1)$ refers to the dependence on $N$, while $R_n$ is integrable on the domain in question.
In \cite{Fo99b} only the leading term in (\ref{G3a}) and (\ref{G3b}) was computed, so in particular
(\ref{b1a}) and (\ref{b1b}) were not rigorously established. This ``technical issue'', essentially
asking for uniform estimates of the infinite sums (\ref{SMP}) and (\ref{SN1}), is the reason
that \cite{Fo99b} was not submitted for publication.

\subsection{Special superimposed ensembles}
With the task of providing uniform asymptotics of (\ref{SMP}) and (\ref{SN1}) being
technically difficult, necessity dictates seeking an alternative strategy. For this one should bring
to the fore the notion of universality, which tells us (for example) that there is nothing
canonical about the finite $N$ ensemble (\ref{qA3a}) in regard to studying the limiting
distribution odd(OEsoft)${}^2$. 
Thus for a general weight function $g_1(x)$
\begin{eqnarray}
&&
E_1(n;(s,\infty);{\rm odd}({\rm OE}_N(g_1(x)) \cup {\rm OE}_N(g_1(x))) = \sum_{l=0}^{2n} E_1(2l-1;(s,\infty);g_1(x)) \nonumber \\
&& \qquad \times  \Big (
E_1(l;(s,\infty);g_1(x)) + E_1(l-1;(s,\infty);g_1(x)) \Big )
\end{eqnarray}
which follows immediately from the definition of the superimposed ensemble. Hence, the universality
of the gap probability in the superimposed ensemble is a consequence of the universality at
the edge of the ensemble OE${}_N(g_1(x))$, which is known from \cite{DG05}.
In particular, instead of considering the soft edge scaling of the
superposition of Gaussian orthogonal ensembles, we may just as well consider the soft edge
scaling of superimposed Laguerre orthogonal ensembles
\begin{equation}\label{AS}
{\rm odd} \Big ( {\rm OE}_N(x^a e^{-x/2}) \cup  {\rm OE}_N(x^a e^{-x/2}) \Big ),
\end{equation}
and more particularly this for any one value of $a$. So the question now is,
amongst the ensembles (\ref{AS}) is there a value of the parameter $a$
for which the correlations have an
explicit form which is easier to analyze that that in (\ref{SN})?

That there is a special ensemble amongst (\ref{AS}) is seen from a theorem in \cite{FR01}.
This theorem classifies all continuous weight functions $g_1, g_2$ such that
\begin{equation}\label{AS1}
{\rm even} \Big ( {\rm OE}_n(g_1) \cup  {\rm OE}_n(g_1) \Big ) = {\rm UE}_n (g_2)
\end{equation}
where UE${}_n(g_2)$ refers to (\ref{2}) with $\beta = 2$ (here the U denotes unitary symmetry
and even refers to the labelling of the eigenvalues countered from the right,
which is the soft edge). In fact up to a fractional linear transformation 
there are only two weight functions with this property,
\begin{equation}\label{AS2}
(g_1, g_2) = \left \{
\begin{array}{ll} (e^{-x/2}, e^{-x}), & x>0 \\
((1-x)^{(a-1)/2}, (1-x)^a ), & -1 < x < 1.
\end{array} \right.
\end{equation}
Hence of the superimposed Laguerre ensembles in (\ref{AS}), the case $a=0$ is distinguished
by the property (\ref{AS1}). In keeping with this special feature, for the correlations of
every odd labelled eigenvalue as required by (\ref{AS}), it allows
the structured formula  \cite{FR02}
\begin{equation}\label{JB}
\rho_{(n)}^{{\rm odd(LOE}{}_N^0)^2}(x_1,\dots,x_n) =
\det \Big [ - {\partial \over \partial x_j} \int_0^{x_k} K_N(x_j,u) \, du
\Big ]_{j,k=1,\dots,n}
\end{equation}
where $K_N$ refers to (\ref{SS2}) with $g_2(x) = e^{-x}$, $p_j(x) = L_j^0(x)$
(here $L_j^0(x)$ denotes the Laguerre polynomials with parameter $a=0$). 

One sees immediately that the structure exhibited in 
(\ref{JB}) gives a much cleaner expression than that exhibited by (\ref{SN}). The task is to
compute uniform asymptotics of this under soft edge scaling, which for the $a=0$ Laguerre
ensemble is obtained by replacing coordinates \cite{Fo93a}
$
x \mapsto  4N + 2 (2N)^{1/3} X
$ then taking the limit $N \to \infty$.
It is a straightforward  exercise using the uniform estimate \cite{Ol74}
\begin{equation}\label{JBa}
e^{-x/2} L_N^0(x) = {(-1)^N \over (2 N)^{1/3} } {\rm Ai}(t) + O(e^{-t}) o(N^{-1/3})
\end{equation}
where $x = 4N + 2(2N)^{1/3} t$, valid for $t \in [t_0, \infty)$ to obtain the uniform asymptotic
expansion
\begin{equation}\label{JB1}
2 (2N)^{1/3} K_N(4N +  (2N)^{1/3} s, 4N + (2N)^{1/3} t) =
K^{\rm soft}(s,t) + O(e^{-(s+t)}) O(N^{-1/3})
\end{equation}
valid for $t,s \in [t_0,\infty)$. Further, this expansion can be differentiated with
respect to $t$ or $s$.  However, as written in (\ref{JB})
the argument $u$ in $K_N(x_j,u)$ takes on
values which are to leading order in $[0,4N]$ instead of $[4N,\infty)$ as in (\ref{JB1}).
As noted in \cite[Lemma 13]{FR02}, this can be circumvented by using the identity
$$
\int_0^x K_N(y,u) \, dy = (-1)^{N-1} \int_y^\infty e^{-u/2} {d \over d u} L_N^0(u) \, du
- \int_x^\infty K_N(y,u) \, du.
$$
Use of (\ref{JBa}) and (\ref{JB1}) then gives the uniform asymptotic expansion
\begin{eqnarray}
&&
2 (2N)^{1/3} \int_0^{4N + 2(2N)^{1/3} X} K_N(4N+2 (2N)^{1/3} Y, u) \, du \nonumber \\
&& \qquad = \int_Y^\infty {\rm Ai}(u) \, du -
 \int_X^\infty K^{\rm soft}(Y,u) \, du + O(e^{-(X+Y)}) O(N^{-1/3}) ,
\end{eqnarray}
which furthermore remains valid upon differentiating with respect to $X$ or $Y$. Noting
from the integral form of the kernel (\ref{1.8a}),
$$
K^{\rm soft}(x,y) = \int_0^\infty {\rm Ai}(x+u) {\rm Ai}(y+u) \, du
$$
that
$$
- {\partial \over \partial Y} \int_{-\infty}^X K^{\rm soft}(Y,u) \, du =
K^{\rm soft}(X,Y) + {\rm Ai}(Y) \int_{-\infty}^X {\rm Ai}(t) \, dt,
$$
and substituting this in (\ref{JB1}) then substituting the result in (\ref{JB}) we deduce
\begin{eqnarray}\label{2.19}
&& (2(2N)^{1/3})^n \rho_{(n)}^{{\rm odd(LOE}{}_N^0)^2}(4N+2(2N)^{1/3} x_1, \dots,
4N+2(2N)^{1/3} x_N) \nonumber \\
&& \qquad =  \rho_{(n)}^{{\rm odd(OEsoft)}{}^2}(x_1,\dots,x_n) + 
e^{-(x_1+\cdots + x_n)}O(N^{-1/3})
\end{eqnarray}
where $ \rho_{(n)}^{{\rm odd(OEsoft)}{}^2} $ is specified by (\ref{2.2a}). It is immediate from
this that the analogue of (\ref{G1}) holds true. But according to \cite{So01} the latter is
sufficient for the validity of the limit formula
\begin{equation}\label{F1}
\lim_{N \to \infty} E_N^{{\rm odd(LOE}{}_N^0)^2}((4N+2(2N)^{1/3} s, \infty);\xi) =
E^{{\rm odd(OEsoft)}{}^2}((s,\infty);\xi),
\end{equation}
with $E^{{\rm odd(OEsoft)}{}^2}((s,\infty);\xi)$ specified by (\ref{b1a}). Hence, for the
soft edge scaling, our sought identity has been established, providing us with a rigorous
justification of the identity (\ref{qA2}).

It remains to establish a similar limit formula for the hard edge. Of the two special pairs of
weights (\ref{AS2}), the second pair near $x=1$ exhibits a general hard edge singularity.
Recalling that the ensemble OE${}_N((1-x)^a(1+x)^b)$ with $-1 < x < 1$ is referred to as the
Jacobi orthogonal ensemble JOE${}_N^{a,b}$, the left hand side of (\ref{AS1}) for the second pair in
(\ref{AS2}) refers to the even labelled eigenvalues counted from the right in the
ensemble JOE${}_N^{(a-1)/2,0} \cup {\rm JOE}{}_N^{(a-1)/2,0}$. Analogous to (\ref{JB1}) the
$n$-point correlation function for the odd labelled eigenvalues of this ensemble is given by
the structured formula \cite[eq.~(2.16)]{FR02}
\begin{equation}\label{JPT}
\rho_{(n)}^{ {\rm odd(JOE}{}_N^{(a-1)/2,0})^2} (x_1,\dots,x_n) =
\det \Big [ - {\partial \over \partial x_j} (1 - x_j) \int_{-1}^{x_k} (1 - x_j)
\tilde{K}_N(x_j,u) \, du \Big ]_{j,k=1,\dots,n}.
\end{equation}
 Here $\tilde{K}_N(x,y) = K_N(x,y) / ((1-x)(1-y))^{1/2}$ where, with $P^{(a,b)}(x)$ denoting the
Jacobi polynomials,  $K_N(x,y)$ refers to (\ref{SS2}) with $g_2(x) = (1-x)^a$ and
$p_j(x) = P_j^{(a,0)}(x)$.

In (\ref{JPT}) the integration variable $u$ is not confined to the neighbourhood of the hard
edge. To avoid this potential problem, we make use of the identity \cite[eq.~(3.69)]{FR02}
\begin{equation}\label{TX}
(1 - x) \int_{-1}^1 \tilde{K}_N(x,u) \, du = - 2 \int_x^1 \tilde{K}_N(-1,u) \, du
\end{equation}
where we regard $\tilde{K}_N(-1,u)$ specified by the final form in (\ref{SS2}), supplemented by
the special value $P_j^{(a,0)}(-1) = (-1)^j$.

The hard edge scaling limit in the neighbourhood of $x=1$ requires replacing the coordinates
$x \mapsto 1 - X/2 N^2$ then taking $N \to \infty$. To analyze the latter limit, we make use
of the uniform asymptotic expansion \cite{Sz75}
\begin{eqnarray}
&& \Big ( \sin {\theta \over 2} \Big )^a \Big ( \cos {\theta \over 2} \Big )^b
P_n^{(a,b)}(\cos \theta) \nonumber \\ && \quad
=
n^{-a} {\Gamma(n+a+1) \over n!}
\sqrt{ {\theta \over \sin \theta} } J_a((n+(a+b+1)/2) \theta) + \theta^{a+2} O(n^a)\label{XN1}
\end{eqnarray}
valid for $0 < \theta < c/n$, $(c > 0)$, which furthermore remains valid upon differentiation
with respect to $\theta$. With use made too of (\ref{TX}), this gives
\begin{equation}\label{XN2}
(1 - x) \int_{-1}^y \tilde{K}_N(x,u) \, du \Big |_{x = 1 - X/2N^2 \atop
y = 1 - Y/2N^2} = - X^{1/2} \int_Y^\infty v^{-1/2} K^{\rm hard}(X,v) \, dv +
O\Big ({1 \over N} \Big ) O(1),
\end{equation}
which furthermore remains valid upon differentiation, with the dependance on $X,Y$ again $O(1)$.
Using the integral form of the kernel (\ref{qA4}),
$$
K^{\rm hard}(x,y) = {1 \over 4} \int_0^1 J_a(\sqrt{xt}) J_a(\sqrt{yt}) \, dt
$$
we have that
$$
{\partial \over \partial X} X^{1/2} \int_Y^\infty v^{-1/2} K^{\rm hard}(X,v) \, dv =
K^{\rm hard}(X,y) + {J_a(\sqrt{X}) \over 2 \sqrt{Y} }
\int_{\sqrt{Y}}^\infty J_a(t) \, dt.
$$
Substituting this in (\ref{XN2}), substituting the result in (\ref{JPT}), and recalling
(\ref{2.3a}) we conclude
\begin{eqnarray}\label{2.19a}
&& \Big ( {1 \over 2 N^2} \Big )^n \rho_{(n)}^{ {\rm odd(JOE}{}_N^{(a-1)/2,0})^2} (
1 -  x_1/2N^2, \dots, 1 - x_n/2N^2) 
\nonumber \\
&& \qquad =  \rho_{(n)}^{{\rm odd(OEhard)}{}^2}(x_1,\dots,x_n;(a-1)/2) + O \Big ( {1 \over N}
\Big ) O(1) 
\end{eqnarray}
where $ \rho_{(n)}^{{\rm odd(OEhard)}{}^2} $ is specified by (\ref{2.3a}).
From this uniform asymptotic expansion 
it is immediate 
that the analogue of (\ref{G2}) holds true, and again by appealing to \cite{So01} the latter is
sufficient for the validity of the limit formula
\begin{equation}\label{F2}
\lim_{N \to \infty} E_N^{{\rm odd(JOE}{}_N^{(a-1)/2,0})^2}((1-s/2N^2,1);\xi) =
E^{{\rm odd(OEhard)}{}^2}((0,s);\xi;(a-1)/2),
\end{equation}
with $E^{{\rm odd(OEhard)}{}^2}$ specified by (\ref{b1b}). This is the sought limit
formula for the hard edge scaling, complementing  (\ref{F1}) for the soft edge scaling,
and providing us with a rigorous justification of (\ref{qA3}).

\end{document}